\documentclass[twocolumn,showpacs,preprintnumbers,amsmath,amssymb,prl]{revtex4}
\usepackage{graphicx}
\usepackage{dcolumn}
\usepackage{bm}
\usepackage{trfsigns}
\usepackage{sidecap}
 
\usepackage{xcolor}
 
\def\ket#1{\lvert#1\rangle}
\def\bra#1{\langle #1\lvert}
\def\braket#1#2{\langle #1\lvert#2\rangle}

\def\i{\mathrm{i}}

\begin{document}
 
%\preprint{APS/123-QED}

\title{General complex Spin Weak Values  obtained in Matter-Wave Interferometer Experiments}

\author{Stephan Sponar$^1$}
\email{sponar@ati.ac.at}
\author{Tobias Denkmayr$^1$}
\email{tdenkmayr@ati.ac.at}
\author{Hermann Geppert$^1$}
\author{Hartmut Lemmel$^{1,2}$}
\author{Alexandre Matzkin$^3$}
\author{Yuji Hasegawa$^1$}
\email{hasegawa@ati.ac.at}

\affiliation{%
$^1$Atominstitut, Vienna University of Technology Stadionallee 2, 1020 Vienna, Austria\\
$^2$Institut Laue-Langevin, 6, Rue Jules Horowitz, 38042 Grenoble Cedex 9, France\\
$^3$ Laboratoire de Physique Th$\acute{e}$orique et Mod$\acute{e}$lisation, CNRS Unit$\acute{e}$ 8089, Universit$\acute{e}$ de Cergy-Pontoise, 95302 Cergy-Pontoise cedex, France}

\date{\today}
\begin{abstract}
Weak values of the spin operator $\hat S_z$ of massive particles, more precisely neutrons, have been experimentally determined by applying a novel measurement scheme. This is achieved by coupling the neutron's spin weakly to its spatial degree of freedom in a single-neutron interferometer setup. The real and imaginary parts as well as the modulus of the weak value are obtained by a systematical variation of pre- and post-selected ensembles, which enables to study the complex properties of  spin weak values. 
\end{abstract}


\maketitle

The meaning of weak values has been an issue of heated debates ever since the concept was introduced by Aharonov, Albert and Vaidman (AAV) in 1988  \cite{Aharonov88}. Unlike in the von Neumann measurement approach, where the outcome of the measurement is an eigenvalue of the observable and the premeasurement state collapses into the corresponding eigenstate  \cite{Neuman32}, the obtained values, so called {\em{weak values}}, may lie far outside the range of the observable's eigenvalues. A weak measurement of an observable $\hat A$ invokes three steps:  (i) preparation of an initial quantum state $\vert\psi_{\rm i}\rangle$ (pre-selection), (ii) a unitary coupling of $\hat A$ with a dynamical variable of a probe system,  via a coupling Hamiltonian, with the coupling being  weak so that the system is minimally disturbed. (iii) post-selection of the final quantum state $ \ket{\psi_{\rm f}}$, by performing a standard projective measurement of another observable $\hat B$ of the quantum system. Finally, the probe system is read out, yielding the weak value given by
\begin{equation}\label{eq:WVdef}
\langle \hat A\rangle _{\rm w}=\frac{\bra{\psi_{\rm f}}\hat A \ket{\psi_{\rm i}}}{\braket{\psi_{\rm f}}{\psi_{\rm i}}}.
\end{equation}

Experimental realizations of the procedure proposed by AAV have been performed  using various optical setups \cite{Ritchie91,Suter95,Hickmann04,Gisin04,Wiseman05, Hosten08,Hofmann11}. More recently experiments addressing average trajectories of single photons in a two-slit interferometer \cite{Steinberg11}, direct measurement of the quantum wavefunction \cite{Lundeen11}, or violation of Heisenberg's measurement-disturbance relationship \cite{Steinberg12,Ringbauer14,Kaneda14} have been performed. Measurements of weak values in at least partially non-photonic schemes are extremely recent, involving transmons in superconducting circuits \cite{Groen13}  and spontaneous emission of photons from atoms  \cite{Shomroni13}. Moreover, most experiments have either measured purely real or purely imaginary weak values, not complex quantities. Quantum paradoxes such as the three-box problem \cite{Steinberg04}, Hardy's  paradox \cite{Steinberg09,Yokota09} or the quantum cheshire cat \cite{Aharonov13,Denkmayr13} have been demonstrated by using weak values.

\begin{figure*}
\centering
 \includegraphics[width=1\textwidth]{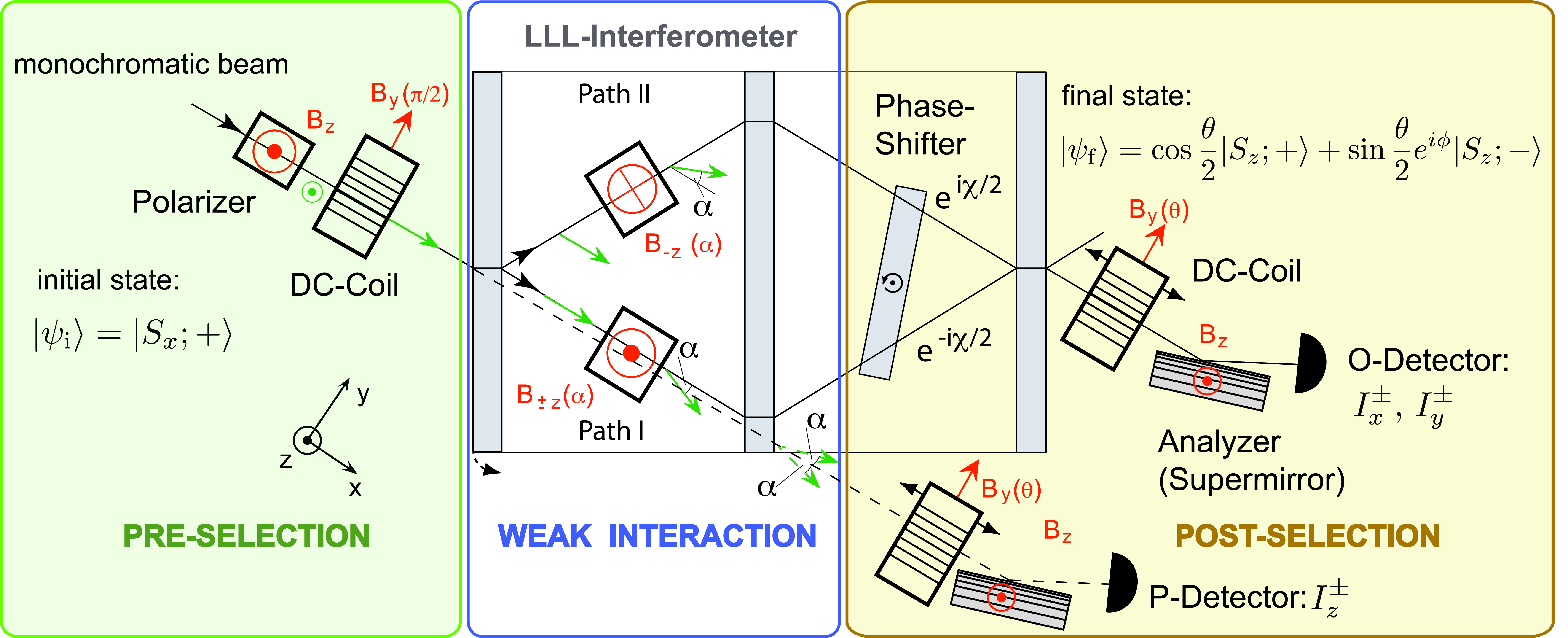} \caption {Schemetic illustration of  triple Laue (LLL) neutron interferometer experiment for a weak measurement of the Pauli spin operator $\hat\sigma_z^{\rm s}$. The setup consists of three stages: (i) pre-selection (green); using a polariser (initial polarisation is along $=z$ direction) and a $\pi/2$-spinrotator the initial state $\ket{\psi_{\rm i}} =\ket{S_x;+}$ is prepared. (ii) weak interaction (blue); in the interferometer  a weak spin rotation by $\pm\, \alpha$ is applied in arm I and II, respectively. (iii) post-selection (brown); a combination of a spin-rotator and a analysing supermirrow is used to post-select on the final state $ \ket{\psi_{\rm f}(\theta,\phi)}=\cos({\theta}/{2})\ket{S_z;+}+\sin({\theta}/{2})e^{\i \phi}\ket{S_z;-}$, before count rate detection in the O-detector. The additional P-detector is required for the determination of the imaginary part of $\langle\hat\sigma_z^{\rm s}\rangle_{\rm w}$, where no interference effects are  nessecarry.\label{fig:IFMsetup}}
\end{figure*}
 
In this paper we report the full determination of the weak value of the Pauli spin operator $\hat\sigma_z$, i.e., its real and imaginary part, as well as its absolute value  (for simplicity $\hat\sigma_z$ is the relevant observable rather than $\hat S_z=\hbar/2\,\hat\sigma_z$). The real and imaginary contributions of the weak value of  $\hat\sigma_z$ are measured using matter wave interferometry with neutrons. Our experiment illuminates a peculiarity of the weak value, namely that it is a complex number in general. Purely imaginary weak values have been utilized to observe amplification effects in weak-measurement-based quantum metrology \cite{Hosten08,Dixon09,Feizpour11}. However, the physical account of the imaginary part of the weak value is still under active discussion \cite{Steinberg95,Jozsa07,Dressel12}. Therefore a systematic investigation of both, the real and the imaginary component of the weak value is of high importance. 

The underlying idea of our  neutron optical experiment is to couple the spin component weakly to another degree of freedom (acting as probe system) which is also a two-level system, rather than a continuous variable system, as proposed in  \cite{Aharonov88}. The probe system applied in our neutron optical experiment are the two paths of a triple Laue neutron interferometer ($\ket{\rm I}$ and $\ket{\rm II}$), which is depicted in Fig.\,\ref{fig:IFMsetup}. 

Our experimental apparatus has several advantages compared to the original setup proposed by AAV in \cite{Aharonov88}: First, in conventional neutron beam experiments the fact that the beam-size (typically at the order of a few cm) is much larger than the coherent-length of the neutron ($\mu$m range), causes problems for weak measurements. Second, instead of measuring the probability distribution of a continuous variable we just have to measure a pair of intensities for each component of the weak value. 
Qubit-qubit weak measurements have already been realized in single photon experiments \cite{Baek08,lund10,Steinberg12}. 
 
Neutron interferometry has established a powerful tool for investigation of fundamental quantum mechanical concepts such as the 4$\pi$ spinor symmetry of fermions, spin- superposition law, and various gravitational effects \cite{RauchBook}, in addition to topological phases \cite{Wagh97,Hasegawa05Cyc,Filipp05NonC}. Entanglement between different degrees of freedom (the neutron's spin and path) has been verified by violation of a Bell inequality \cite{Hasegawa03,Sponar10}. In addition, preparation of different types of genuine tripartide entanglement \cite{Hasegawa10,Erdoesi13},  as well as demonstration of the contextual nature of quantum mechanics \cite{Hasegawa2006contextual,Bartosik09} have been performed successfully. In the present experiment we make use of a degree of freedom of a single particle system whose spin is weakly measured, as suggested by AAV in  \cite{Aharonov88}. Within the  perfect silicon-crystal interferometer a beam of single neutrons is split by amplitude division and superposed coherently after passing through different regions of space, resulting in an interference effect of matter waves \cite{Rauch74}. During these space-like separation (typically a few centimeters)  the neutron's wavefunction aquires phase shifts due to nuclear and magnetic interaction. The former is applied to induce an adjustable phase factor $e^{\i\chi}$ ($\chi=N_{\rm ps}b_{\rm c}\lambda D$ with the thickness of the phase shifter plate $D$, the neutron wavelength $\lambda$, the coherent scattering length $b_{\rm c}$ and the particle density $N_{\rm ps}$ in the phase shifter plate). By rotating the plate $\chi$ can be varied systematically, due to the change of the relative optical path length. This yields the well known intensity oscillations of the two beams emerging behind the interferometer, usually denoted as O- and H-beam. 

In our experimental procedure the initial spin state is given by 
\begin{equation}\label{eq:initialState}
\ket{\psi_{\rm i}}=\ket{S_x;+},
\end{equation}
completing the pre-selection of the spin. Inside the interferometer the incident wave function $\ket{\Psi_{\rm inc}}$ is denoted as 
\begin{equation}\label{eq:initialTotalState}
\ket{\Psi_{\rm inc}}=\frac{1}{\sqrt2}\left(\mathrm e^{\i\chi}\ket{\rm I}+  \ket{\rm II}\right)\ket{S_x;+}.
\end{equation}
The weak measurement  is done by inducing small path dependent spin rotations; in each arm, a magnetic field $\vec B$ along $z$ but pointing in opposite directions is applied. The interaction Hamiltonian between the spin and the path dependent magnetic field can be written as
\begin{equation}\label{eq:Hint}
 \hat{H}_{\rm int}=-{\mu}\sigma_z^{\rm s}({B_z}\hat\Pi_{\rm I}-{B_z}\hat\Pi_{\rm II})= -\mu B_z\hat\sigma_z^{\rm s}\hat\sigma_z^{\rm p},
\end{equation}
where ${\mu}$ is the neutron's magnetic moment and $B_z$ the applied magnetic field pointing in $z$-direction and $\hat\sigma_z^{\rm s}$ is the Pauli spin operator. $\hat\Pi_{\rm I}$ and $\hat\Pi_{\rm II}$ are projection operators to path I and II, respectively and can be expressed using Pauli matrix formalism as ${\hat\sigma_z^{\rm p}=\hat\Pi_{\rm I}-\hat\Pi_{\rm II}}$. After the weak interaction the wave function evolves as 
\begin{equation} 
\ket{\Psi'}=e^{-{\rm i}/\hbar\int{\rm d}t\hat{H}_{\rm int}}\ket{\Psi_{\rm inc}}\thickapprox\left(1-\frac{{\rm i} \alpha \hat\sigma_z^{\rm s}\hat\sigma_z^{\rm p}}{2}\right)\ket{\Psi_{\rm inc}},
\end{equation}
where the parameter $\alpha$ defines the angle of spin rotation. Since $\alpha$ is proportional to the expression $\mu B_z$ from the interaction Hamiltonian in Eq.(\ref{eq:Hint}), $\alpha$ accounts for the interaction strength of the weak measurement. The post-selection onto an arbitrary spin state  
\begin{equation}\label{eq:final}
 \ket{\psi_{\rm f}(\theta,\phi)}=\cos\frac{\theta}{2}\ket{S_z;+}+\sin\frac{\theta}{2}e^{\i \phi}\ket{S_z;-},
\end{equation}
with polar angle $\theta$ and azimuthal angle $\phi$. After re-exponentiation the final wave function reads
\begin{equation}\label{eq:wavefunctionweakspin}
 \ket{\Psi_{\rm fin}}=\braket {\psi_{\rm f}}{\psi_{\rm i}}\frac{1}{\sqrt 2}\left(\mathrm e^{\i\chi}e^{-\i\alpha\langle\hat{\sigma}_z^{\rm s}\rangle_{\rm w}/2} \ket{\rm I}+\mathrm e^{\i\alpha\langle\hat{\sigma}_z^{\rm s}\rangle_{\rm w}/2} \ket{\rm II}\right)\ket{\psi_{\rm f}},
\end{equation}
with the spin weak value $\langle\hat{\sigma}_z^{\rm s}\rangle_{\rm w}=\bra{\psi_{\rm f}}\hat \sigma_z^{\rm s} \ket{\psi_{\rm i}}/\braket{\psi_{\rm f}}{\psi_{\rm i}}$. Here the real part of the weak value of the spin operator $\hat{\sigma}_z^{\rm s}$ acts as an additional phase in the wave function, while the imaginary part affects the amplitudes. The real and imaginary component, and the modulus of the weak value can be experimentally determined from 
 \begin{equation}\label{eq:wvreal}
\Re\langle \hat{\sigma}_z^{\rm s}\rangle_{\rm w}=\frac{1}{\alpha}\arcsin\big(\bra{\Psi_{\rm fin}}\hat{\sigma}_y^{\rm p}\ket{\Psi_{\rm {fin}}}\big)=\frac{1}{\alpha}\arcsin\left(\frac{I_y^+-I_y^-}{I_y^++I_y^-} \right),
\end{equation}
\begin{equation}\label{eq:wvimag}
\Im\langle \hat{\sigma}_z^{\rm s}\rangle_{\rm w}=\frac{1}{\alpha}{\rm atanh}\big(\bra{\Psi_{\rm fin}}\hat{\sigma}_z^{\rm p}\ket{\Psi_{\rm {fin}}}\big)=\frac{1}{\alpha}{\rm atanh}\left(\frac{I_z^+-I_z^-}{I_z^++I_z^-} \right),
\end{equation}
and
\begin{equation}\label{eq:wvabs}
\vert \langle \hat{\sigma}_z^{\rm s}\rangle_{\rm w}\vert=\frac{1}{\alpha}\arccos\big(\bra{\Psi_{\rm fin}}\hat{\sigma}_x^{\rm p}\ket{\Psi_{\rm {fin}}}\big)=\frac{1}{\alpha}\arccos\left(\frac{I_x^+-I_x^-}{I_x^++I_x^-} \right),
\end{equation}
respectively. Here $I_{x,y,z}^\pm$ denote the measured intensities obtained by employing different phase shift values or path recombinations, thereby defining the respective measurement direction of the path two-level system.

The experiment was carried out at the neutron interferometer instrument S18 at the high-flux reactor of the Institute Laue-Langevin (ILL) in Grenoble, France. A schematic illustration of the setup is depicted in Fig.\,\ref{fig:IFMsetup}. The actual neutron optical experiment consist of three stages: (i) pre-selection, (ii) weak interaction and (iii) post-selection (see Fig.\,\ref{fig:IFMsetup}). 

(i) Pre-selection: A monochromatic beam with mean wavelength $\lambda_0=1.91\,\AA$ ($\lambda/\lambda_0\sim 0.02$) and $5\,\times\,5\,\rm mm^2$ beam cross section is polarized in $z$-direction. Before the neutrons enter the interferometer the spin is rotated into the $x$ direction by a $\pi/2$ spin-turner. The spin turner consists of a DC coil with its field $B_y$ pointing in $y$ direction, where due to Larmor precession within DC coils the spin precesses about the $y$-axis. $B_y$ is chosen such that it induces a $\pi/2$ spin rotation, thereby preparing the initial spin state $\ket{\psi_{\rm i}}=\ket{S_x;+}$. Behind the first plate of the interferometer (beam splitter), the neutron's wave function is found in a coherent superposition of the two sub beams belonging to the right ($\ket{\rm I}$) and the left path($\ket{\rm II}$), which are laterally separated by several centimeters.
\begin{figure}[t]
\centering
\includegraphics[width=0.45\textwidth]{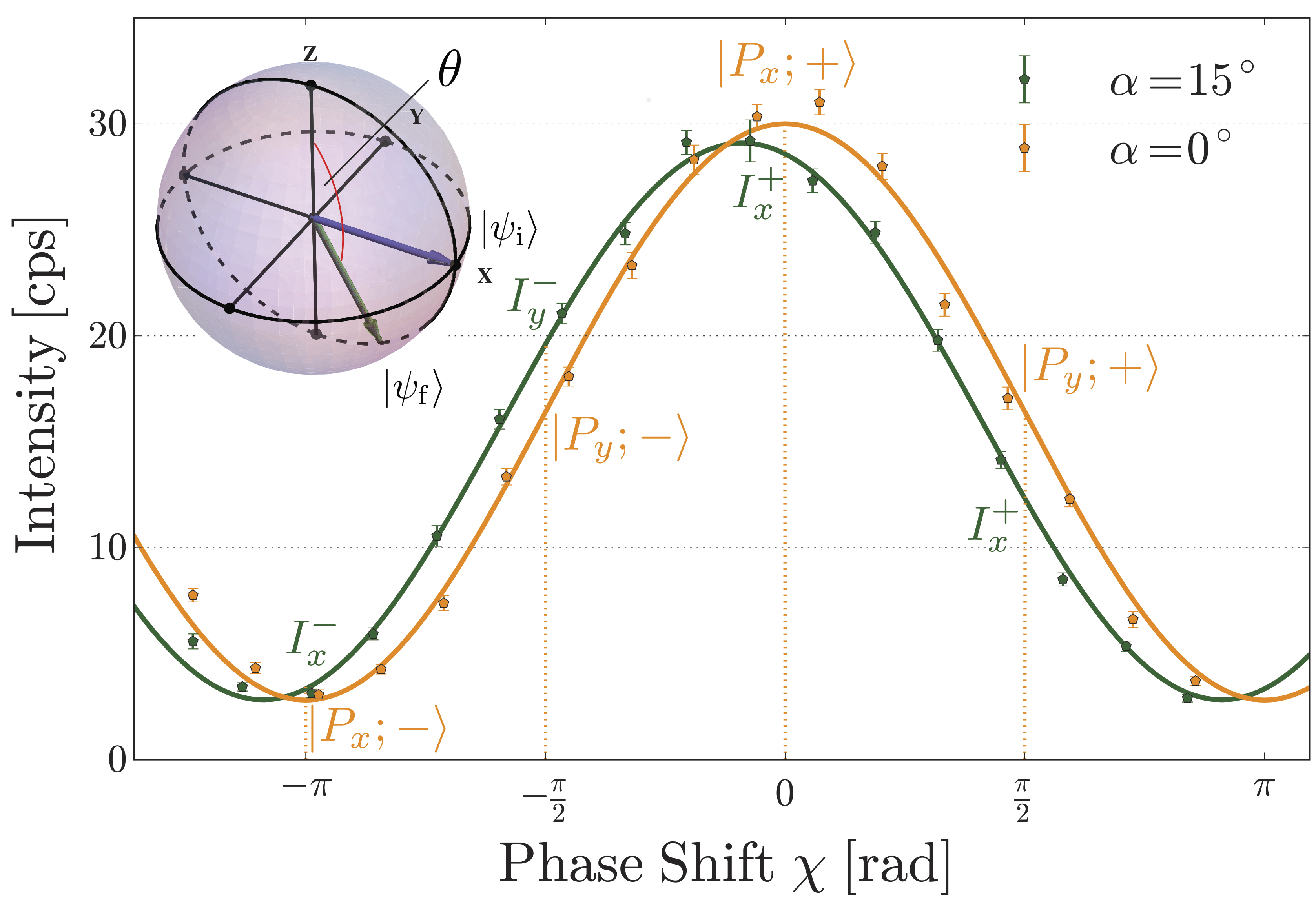}
\caption {Typical interference pattern for $\theta=5\pi/6$ and $\phi=0$ recorded in O-detector without (orange) and with weak spin rotations (green). From the former the phase shifter positions for the states $\ket{P_{x};\pm}$ and $\ket{P_{y};\pm}$ (dashed lines) are obtained and the actual Intensities $I_x^{\pm}$ and $I_y^{\pm}$ are determined from the latter (here the notation $\ket{P_x;\pm}=1/\sqrt 2\left(\ket{\rm I}\pm\ket{\rm II} \right)$ and $\ket{P_y;\pm}=1/\sqrt 2\left({\ket{\rm I}}\pm\i\ket{\rm II} \right)$ is used). \label{fig:Counts}} 
\end{figure}

(ii) Weak interaction: Small spin rotations of $\pm\,\alpha$ are introduced by accelerator coils. These coils are aligned in Helmholz configuration and create a local modification of the static overall guide field $B_z$. Thereby the Larmor frequency is increased in path $\rm I$ and decreased in path $\rm II$, leading to the different spin rotations of $\pm\,\alpha$ respectively.  

\begin{SCfigure*}
\centering
\includegraphics[width=0.835\textwidth]{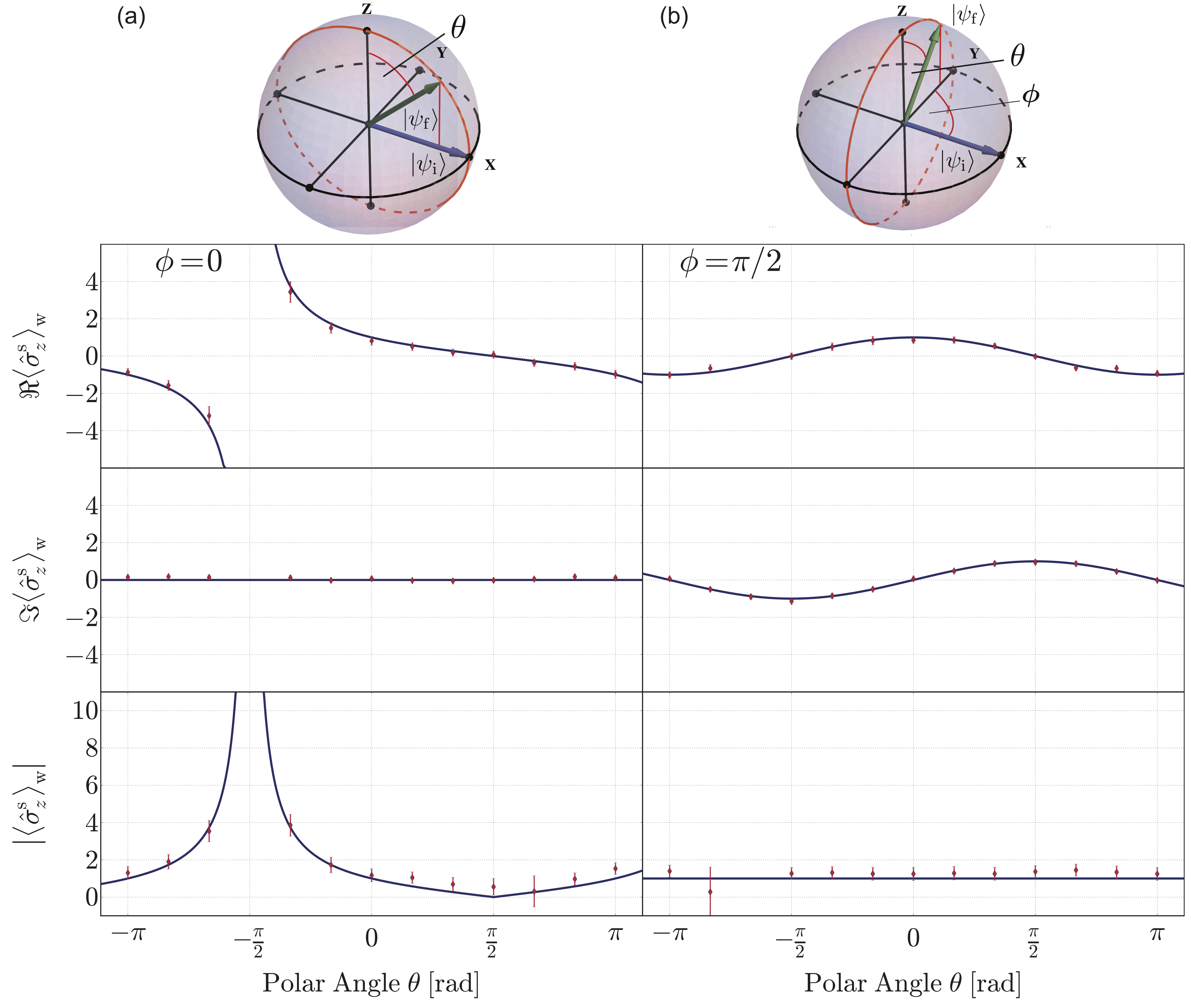}
\caption {The Real and imaginary component as well as the modulus of the weak value of $\hat\sigma^{\rm s}_z$ with the spin pre-selected in state $\ket{\psi_{\rm i}}=\ket{S_x;+}$ and post-selected to $\ket{\psi_{\rm f}(\theta,\phi)}$ are displayed as a function of the polar angle $\phi$ for two selected azimuthal angles $\phi$. (a) For $\phi=0$ the imaginary part of $\langle\hat{\sigma}_z^{\rm s}\rangle_{\rm w}$  is zero and the real part of $\langle\hat{\sigma}_z^{\rm s}\rangle_{\rm w}$, exhibits values which lie outside the usual range of spin eigenvalues, i.e., $\pm$ 1. (b) For $\theta=\pi/2$ real and imaginary component of $\langle\hat{\sigma}_z^{\rm s}\rangle_{\rm w}$ oscillate in quadrature yielding a constant value of $\vert\langle\hat{\sigma}_z^{\rm s}\rangle_{\rm w}\vert$ irrespective of the polar angle of the post-selected state.\label{fig:Results}}
\end{SCfigure*}

(iii) Post-selection:  The spin is rotated by a polar angle $\theta$ with a static field spin turner mounted on a translation stage whose position gives the azimuthal angle $\phi$, due to Larmor precession in the static magnetic guide field. With this combination an arbitrary final state $ \ket{\psi_{\rm f}(\theta,\phi)}=\cos\frac{\theta}{2}\ket{S_z;+}+\sin\frac{\theta}{2}e^{\i \phi}\ket{S_z;-}$ can be post-selected. The spin is finally analyzed by a spin- dependent reflection from bent Co-Ti supermirror array.

For the read out of the probe system, measurements of the Pauli operators $\hat\sigma^{\rm p}_x$ and $\hat\sigma^{\rm p}_y$ in the path subspace 
are carried out by decomposing them into projection operators:
  \begin{equation}
     \hat\sigma_x^{\rm p}= \hat \Pi^{\rm p}(\chi=0)-\hat \Pi^{\rm p}(\chi=\pi)
  \end{equation}
   and
  \begin{equation}  
    \hat\sigma^{\rm p}_y=\hat \Pi^{\rm p}(\chi=\frac{\pi}{2})-\hat \Pi^{\rm p}(\chi=\frac{3\pi}{2}),
  \end{equation} 
where $\hat\Pi^{\rm p}(\chi)$ is given by projector to the state $({\ket{\rm I}}+e^{\i\chi}\ket{\rm II})/\sqrt 2$. The exact phase shifter positions corresponding to the respective relative  phase shifts $\chi$ are experimentally determined by a reference measurement where the weak spin rotations are not applied ($\alpha=0$), see Fig\,\ref{fig:Counts}. From the  measurement with weak rotations ($\alpha=15\,^\circ$) the count rates of the required intensities $I^{\pm}_x$ and  $I^{\pm}_y$ are recorded in the O-detector. As seen from Eq.(\ref{eq:wvreal}) and Eq.(\ref{eq:wvabs}) real part and modulus of the weak value of the spin operator $\langle\hat{\sigma}_z^{\rm s}\rangle_{\rm w}$ are determined from these intensities.

Concerning the imaginary part of  $\langle\hat{\sigma}_z^{\rm s}\rangle_{\rm w},$  the Pauli matrix $\hat{\sigma}_z^{\rm p}$ of the path two-level system  (which occurs in the interaction Hamiltonian in Eq.(\ref{eq:Hint})) has to be measured. This is a rather alienated experimental configuration: $\hat{\sigma}_z^{\rm p}$ contains the orthogonal path eigenstates $\ket{\rm I}$ and $\ket{\rm II}$, which have to be measured separately and therefore without interference effects. This can be achieved in principle by rotating the interferometer by a few seconds of arc such that the Bragg condition at the first plate of the interferometer is no longer fulfilled and the beam passes the first plate without diffraction. After a weak spin rotation in direction $\pm\,\alpha$, for path $\ket{\rm I}$ and $\ket{\rm II}$ respectively, the post-selection for the final spin state is performed in the same manner as for real and absolute of $\langle\hat{\sigma}_z^{\rm s}\rangle_{\rm w}$. Finally the successfully post-selected neutrons are detected in the P-detector, which can be seen in Fig.\,\ref{fig:IFMsetup} in the lower part of the schematic illustration.  In practice this experimental configuration was realized omitting the interferometer since the interference effects between path I and II are irrelevant. Consequently the measurement was carried out separately  at the polarimeter beam line of the Atominstitut, at the Vienna University of Technology, using the same beam parameters as in the ILL setup.  

The final results are plotted in Fig.\,\ref{fig:Results}. Values of the polar angle $\theta$ are systematically varied in an interval between $-\pi$ and $\pi$ for azimuthal angle $\phi=0$ and   $\phi=\pi/2$, defining the post-selected state  $\ket{\psi_{\rm f}}$. The weak value of the spin operator $\hat\sigma^{\rm s}_z$ is theoretically calculated as
 \begin{equation} \label{eq:WVgeneralkspinstate}
\langle\hat\sigma^{\rm s}_z\rangle_{\rm w}=\frac{\cos\theta}{1+\sin\theta\cos\phi}-\i\frac{\sin\phi\sin\theta}{1+\sin\theta\cos\phi}.
   \end{equation} 
 According to this equation no imaginary contributions of the weak value of $\hat{\sigma}_z^{\rm s}$  are expected for $\phi=0$; the modulus of the weak values equals the modulus of the real part of the weak value. The experimentally determined components of the weak value of $\hat{\sigma}_z^{\rm s}$ are plotted in Fig.\,\ref{fig:Results}\,(a), reproducing the theoretical predictions from Eq.(\ref{eq:WVgeneralkspinstate}), evidently. For $\phi=\pi/2$ real and imaginary component of the weak value of $\hat{\sigma}_z^{\rm s}$ exhibit phase-shifted oscillations but with the same amplitude,  such that a constant value of $\vert\langle\hat{\sigma}_z^{\rm s}\rangle_{\rm w}\vert$ is obtained, which is plotted in Fig.\,\ref{fig:Results}\,(b). 

For the modulus of the weak value, shown in Fig.\,\ref{fig:Results}, the finite contrast of the interferometer had to be taken into account, since in the experiment an average contrast of $\sim80\,\%$ was achieved. In addition, a background subtraction was performed. For the real part of the spin operator's weak value it was not necessary to normalize the measurement data on the contrast, only the background was subtracted to obtain the real part of the weak value values. For the operator's real part, spin weak values ranging from -3.2 to 3.4 were measured, which is clearly outside the eigenvalue spectrum of the spin operator $\hat\sigma_z^{\rm s}$. Note that for $\theta=\pi/2$ and $\phi=0$ initial and final state coincide. Thus the weak value reduces to the expectation value $\langle S_x;+\vert\hat\sigma_z^{\rm s}\vert S_x;+\rangle$ which yields zero, as expected. 

In summary, we performed a complete determination of the weak value of the Pauli spin operator $\hat \sigma_z$, extracting the weak value's modulus, its real and imaginary part with high accuracy from the raw data. It has to be stressed that the results are a purely quantum mechanical effect, since the neutron is described in terms of matter waves: hence no classical theory can be applied to describe the results. Our measurement scheme can be extended to measure arbitrary spin operators weakly, allowing for a state tomography for massive particles via weak measurements \cite{Salvail13}. In addition, our measurement approach is applicable for other two level quantum systems as well, which establishes a new measurement technique that allows to test quantum mechanics at a fundamental level.

We thank Prof. Dipankar Home from the Bose Institute in Calcutta for helpful discussions. This work was financed by the Austrian Science Fund (FWF, Project. Nos. P25795-N02 and P24973-N20) as well as from the Austrian-French binational Amadeus Project (No. FR 06/2012).

\end{document}